# Structural correlation to enhanced magnetodielectric properties of Pr- doped polycrystalline $Gd_{0.55}Pr_{0.45}MnO_3$ at low temperatures

Pooja Pant[1], Harshit Agarwal[1,2] Suresh Bharadwaj[3], Archana Sagdeo[4], M. A. Shaz[1]*

[1] Department of Physics, Institute of Science, Banaras Hindu University, Varanasi-221005

[2] Laboratory of Crystallography, University of Bayreuth, Bayreuth, Germany

[3] UGC DAE Consortium for Scientific Research, Indore

[4] Synchrotrons Utilization Section Raja Ramanna Centre for Advanced Technology Indore-13

**Abstract:**

The effect of improved dielectric properties in the presence of an applied magnetic field is discussed in Pr- doped polycrystalline $Gd_{0.55}Pr_{0.45}MnO_3$ complemented by the structural properties down to 50 K and magnetic properties. Enhanced dielectric permittivity and low dielectric loss represent the strongly field-dependent dielectric behaviour of the sample. The structural characterizations using synchrotron angle dispersive X-ray diffraction confirm the orthorhombic symmetry of the polycrystalline sample with the *Pnma* space group from 300 K to 50 K. There is no structural transition at lower temperatures up to 50 K; however the Mn- O octahedral distortion is reduced. The investigation of dielectric properties for frequencies range 500 Hz to 1 MHz was conducted in the temperature range 8 K – 300 K with and without a magnetic field of 7 Tesla, which shows the high dielectric constant in the 500 Hz frequency region. This confirms the relaxation phenomena in polycrystalline $Gd_{0.55}Pr_{0.45}MnO_3$. *ac* conductivity data shows the increasing trend for frequencies along with the activation energy which increases from low frequencies to higher frequencies. Temperature-dependent *dc* magnetization study shows the negative magnetization in polycrystalline $Gd_{0.55}Pr_{0.45}MnO_3$ at low temperature at 100 Oe applied field, due to spin canting of Mn- Mn magnetic lattice. High coercivity due to competition in between spin ordering of Mn and Gd+Pr magnetic lattice at 5 K are also observed in the field-dependent magnetization study.

**Keywords:** Magnetodielectric, Synchrotron X-ray diffraction, Negative magnetization, Dielectric relaxation

**Corresponding author:** shaz2001in@yahoo.com

## 1. Introduction:

Recent advances in multifunctional device concepts and applications have led to the exploration of multiferroics that possess both ferroelectric and magnetic properties simultaneously. Multiferroics are exotic materials that exhibit coupling between magnetic polarization and ferroelectric ordering. Many practical applications have been developed based on the magnetoelectric effect in multiferroic materials[1–3]. Among the two types of multiferroics, Type- II multiferroics encourage more physics and application because of the intrinsic mutual control of magnetism and ferroelectricity. It can be explained by the fact that the electric polarization 'P' is generated by some specific spin orders including the noncollinear spiral spin (NSS) order and the anti-ferromagnetic (E-AFM) order[4,5]. In rare-earth manganites $RMnO_3$, the structural distortion occurs in the materials due to the mismatch in ionic radius of A and B site cations which correlated to the physical and magnetic properties of the materials. The crystal structure of these types of materials exists in cubic, hexagonal, orthorhombic, or sometimes coexists in two-phase[6]. Structural strength depends on the Mn-O bond length and Mn-O-Mn bond angle, [7,8], surface area, as well as disorder degree, i.e. mismatch A-site cation[9], and size of A cation[10]. The magnetic ordering in such materials is controlled by the Mn-O-Mn bond angle [11,12]. These materials also exhibit relatively strong Jahn- Teller orbital ordering, which is due to Jahn- Teller active $Mn^{3+}$ ions [12]. As a result of the coupling between the rare-earth ions and the Mn subsystem, manganites display anisotropic magnetic and thermodynamic properties at low temperatures[13].

There have also been recent observations of magnetoelectric effects in manganite[14], including hexagonal $RMnO_3$[15]and orthorhombic $RMnO_3$[16]. A room-temperature magnetoelectric and magneto transport study was conducted by Gadani et al.[17] on $LaMnO_3$ manganites. As part of their research to improve spintronics applications, they have also studied electric pulse-induced resistance (EPIR) and magneto EPIR behaviours. Among the base manganites, $GdMnO_3$ and $PrMnO_3$ have gained remarkable attention for their different properties, such as dielectric relaxation, colossal magneto-resistance and magneto-caloric effects around their curie temperatures[18,19]. It is generally accepted that both families are antiferromagnetic materials of the A-type [20,21], however the differences in their phase structures depend on the doping concentration, the sintering temperature, and the synthesis method[22]. Multiferroic $Gd_{1-x}Ho_xMnO_3$ is produced by adding Ho doping, resulting in a transition from the paraelectric phase to the ferroelectric phase. The enhancement of electric polarization is attributed to the transition of the A-type antiferromagnetic state into the spiral spin order due to the doping reducing the angle between Mn-O-Mn bonds[23]. Around its Curie temperatures $T_C$ with X = Ce, Eu, and Y, $Pr_{0.5}X_{0.1}Sr_{0.4}MnO_3$ undergoes a second-order magnetic transition from paramagnetic to ferromagnetic[24].

In recent years, in rare earth metal oxides, magnetization reversal phenomena have received much attention[25,26]. The ferromagnetic materials containing two or more magnetic ions occur in this type of effect or in which one magnetic ion occupies a different crystallographic site. For obtaining the multiferroic in these systems, the promising materials are the double orthomanganite of $R_{1-x}R'_xMnO_3$ (R, R′ = La-Lu, 0<x<0.5), which simultaneous contains two types of cations in rare-earth sublattice. As a result of the parent and doped rare earth metals having the same oxidation state during the substitution, the electroneutrality condition is satisfied in the resultant compositions. Sharma et al. reported in Gd doped $LaCrO_3$ that the magnetization reversal occurs due to the different canted magnetic moments of Gd and Cr elements [27]. This type of magnetic reversal phenomena has also observed in $La_{1-x}Pr_xCrO_3$ [28], $La_{1-x}Gd_xMnO_3$[29], and other manganites [26,30]. The electrical properties of ceramic oxides make them very useful for practical applications. Most studies report the dielectric and ferroelectric behaviour in various $RMnO_3$ manganites[31,32]. $RMnO_3$ manganites exhibit dielectric behaviour because of the A-type antiferromagnetic spin order in which ferromagnetism occurs along the a and, b axis and antiferromagnetism occurs along the c axis. Manganite materials can also control their dielectric properties using external magnetic fields, which is an additional property, i.e. magneto-dielectric properties[33,34]. The relaxation effects of multiferroic manganites have been studied in single crystals, and it has been shown that relaxation effects play an important role in magnetocapacitive behaviour[35].

Recently Pr doped $TbMnO_3$ [36]showed the Zero field-cooled and field-cooled dc magnetization studies at low temperatures down to 5 K, which reveals that magnetic phase transitions vary due to $Pr^{3+}$ substitution at the $Tb^{3+}$ site at low temperatures, a signature of the A-type antiferromagnetic nature of the sample at low temperature, accompanied by ferromagnetic clustering, which motivates us to investigate the structural and magnetic properties of Pr doped $GdMnO_3$. There has been no previous research on Pr doping in $GdMnO_3$ in terms of its magnetodielectric effect on the material. We are expecting a tremendous change in dielectric permittivity in the presence of a magnetic field and anomalous magnetic behaviour of Pr doped GdMnO$_3$ at low temperature. In the present study, Gd$_{0.55}$Pr$_{0.45}$MnO$_3$ (GPMO) has been experimentally investigated for its structure, magnetic, and magneto-dielectric properties.

## 2. Synthesis and Characterization Techniques:

The conventional solid-state synthesis method is used for the synthesis of polycrystalline GPMO at high temperatures. We have taken the high purity oxide precursors such as gadolinium oxide (Gd$_2$O$_3$), manganese oxide (MnO$_2$), and presidium oxide (Pr$_6$O$_{11}$) with stoichiometric proportion to synthesize the sample[37].

Room temperature XRD data was carried out using a PANalytical Emperyon X-Ray Diffractometer having CuKα radiation. Indus-2, a synchrotron radiation source (Raja Ramanna Centre for Advanced Technology, Indore), carried out an XRD experiment on powder using a BL-12 ADXRD beam line. X-rays of 15 keV energy were incident on the target. In the 2θ range of 5–35° with step sizes of 0.01° and a wavelength of 0.81711 Å, the diffraction pattern was recorded with a Huber six-circle diffractometer (model 5020). Using JANA2006 and VESTA software, the refined diffraction pattern was analyzed, and the crystal structure was visualized for the 300 K, 200 K, 100 K, and 50 K temperatures. The Alpha-A frequency analyzer from NOVO-CONTROL was used to measure frequencies from 500 Hz to 1 MHz up to a temperature range of 10 K to 300 K for the dielectric measurement with (7 Tesla external field) and without an external magnetic field was carried out. With SQUID- VSM, a temperature-dependent DC magnetization study was performed at 5 K to 300 K using a 100 Oe magnetic field. A DC magnetization study has also been conducted on polycrystalline GPMO samples at 300 K and 5 K temperatures as a function of the applied magnetic field M (H).

## 3. Results and Discussion:

### 3.1. Structural characterization using lab source X-ray diffraction:

The single phasic polycrystalline nature of Gd$_{0.55}$Pr$_{0.45}$MnO$_3$ has been confirmed by lab source X-ray diffraction to investigate the structural correlation to physical and magnetic properties. **Figure 1(a)** shows the refined X-ray diffraction data of polycrystalline Gd$_{0.55}$Pr$_{0.45}$MnO$_3$ (GPMO) collected at room temperature with wavelength 1.54 Å. It has been seen in the figure that after doping of 45% of Pr at the Gd site, the peak positions are shifted towards the lower 2 theta angle as compared to the parent sample $GdMnO_3$[31]. For purity and proper phase property of the synthesized polycrystalline sample, we have done the Rietveld refinement of the X-ray data using computational software JANA2006. **Figure 1(a)** shows the Rietveld refinement XRD data of the polycrystalline GPMO including the calculated and observed profiles as well as the difference profile. The refined data shows that the synthesized sample is in a single phase and there is no impurity present in it. Also, the sample is crystallizing in an orthorhombic perovskite structure having a *Pnma* space group. Orthorhombic *Pnma* space group is used for indexing the diffraction peaks. The shifting of the peaks in the lower 2 theta angle of GPMO in comparison to $GdMnO_3$ is because of the mismatch in ionic radii of the A-site cation (parent and dopant) because Pr is a higher ionic radii element in comparison to Gd. By the goodness of fit and R-values, we can confirm the excellence of data fitting. The shifting in peak position in the XRD pattern by doping agrees with Vegard's law [38], as the substitution of higher ionic radii Pr at the site of lower ionic radii Gd, leads to the increase in the cell parameters so the lattice structure and crystal structure are slightly distorted as a result. These changes might be the significant parameter for the magnetic properties of the material. The refined orthorhombic crystal structure of GPMO is shown in the inset of **figure 1(a)**, which represents that Pr is occupying the 45% space in the Gd- site properly. The refined lattice parameters and structure parameters are shown in **table 1(a).** From the crystal structure of the sample, we have calculated the bond lengths and bond angle, which suggest the centrosymmetric nature of the structure. **Figures 1(b), (c), (d), and (e)** represent the polyhedral view of the crystal structure, top view of the structure, bond angle, and bond length of the polycrystalline GPMO, respectively. By substituting Pr at the Gd site, the angle between Mn-O-Mn bonds and the length of Mn-O bonds are modified. The detail of the calculated bond length and bond angle values is tabulated in **table 1(b)**. These results indicate that after doping of higher ionic radii element Pr did not introduce phase change in the parent structure of $GdMnO_3$. As we know that in rare-earth manganites, several crystallographic factors show the physical and structural properties of the sample. For this purpose, we have calculated the tolerance factor, mismatch variance size, and the average of A-site cation $<r_A>$ by using ionic radii of the cation and anion. According to Shannon, the ionic radius is calculated[39]. The calculated tolerance factors value is 0.87818, which verifies the orthorhombic distortion condition of the synthesized sample. In the orthorhombic structure, strain induced in the synthesized sample can be calculated by the lattice parameter. For the space group *Pnma*, the orthorhombic strain can be calculated by ε= (a-c)/(a+c), the calculated value of strain is 0.04123. The lattice parameter of the polycrystalline GPMO verified the $c/\sqrt{2} < a < b$ relation, which indicates the sample has characterized the presence of static Jahn Teller distortion [40]. The static Jahn Teller distortion in orthorhombic structure can be evaluated using bond length by the relation,

$$\Delta_{JT} = \frac{1}{N} \sum_{i=1....N} \left[ \frac{(Mn-O)_i - <Mn-O>}{<Mn-O>} \right]^2$$

The calculated value of the J-T distortion in polycrystalline GPMO is 0.00030. It is possible that the low value of JT distortion reflects a change in the one electron bandwidth of the $Mn^{3+}$ ion. The $e_g$ band's one-electron bandwidth is calculated as follows; $w = \frac{\cos(\frac{\omega}{2})}{d^{3.5}_{<Mn-O>}}$, where ω= π- <Mn-O-Mn>. The calculated one electron bandwidth W for polycrystalline GPMO is 0.08354. It is seen that the value of bandwidth is lower which suggests the antiferromagnetic insulating state by localizing the $e_g$ charge carriers. Further study is focused on magneto-dielectric properties of polycrystalline $Gd_{0.55}Pr_{0.45}MnO_3$.

**3.2 Dielectric and Magnetodielectric Study:**

**3.2.1 Dielectric permittivity:**

**Figures 2(a)** shows the dielectric constant (ε′) for the various frequencies between 500 Hz- 1 MHz in the 10 K to 300 K range with and without a magnetic field. From the figures, it has been clear that the different permittivity characteristics existed in the different temperature ranges for 500 Hz to 1 MHz frequency. It is seen from the figure that the dielectric constant is dependent upon the frequency, which indicates the dielectric relaxation phenomena in polycrystalline GPMO. As shown in **figure 2(a),** for the lower frequencies between 500 Hz- 10 kHz effect of temperature in the dielectric constant is minimum at 10 K – 80 K, but after 80 K, ε′ increases up to its maximum value for all the frequencies. The low value of the ε′ at low temperature is mainly due to that at low temperature the electric dipoles cannot orient themselves with the direction of the applied electric field so their contribution is weak to the polarization or dielectric constant. When the temperature is increased, after 80 K, the ε′ increases which mean after a rise in temperature most of the electric dipoles get enough exciting energy to change the external electric field so the ε´ increases rapidly and reached its maximum position for all frequencies. It is seen in the figure that the dielectric constant reached its maximum value for 500 Hz to 10 kHz and then met at one point around 235 K for all the frequencies. For the higher frequencies 50 kHz to 1MHz the dielectric constant is gradually varied from 200 K and after this, it increases rapidly with an increase in temperature. It was seen in **figure 2(a)** that the dispersion of the dielectric constant at low frequencies is greater than that of higher frequencies. When we observed data for low frequency to a higher frequency, the value of ε´ is decreased, this type of behavior can be understood by the principle of polarization mechanisms. In materials, polarizations are divided into four types; dipolar, atomic, electronic, and space charge polarization. All these types of polarizations are frequency-dependent so the sum of these polarizations changes with frequency. So when frequencies are changed then net polarization also changes affects the dielectric permittivity. For low frequencies, all the polarizations take part, so the real part of permittivity has a high value for lower frequencies but when we go through the higher frequencies, the phenomena fade gradually. Also, for the 10 kHz frequency, the effect of temperature in the ε´ is low between 10 K – 150 K but after 150 K, ε´ increases up to its maximum value for 0 Tesla as well as for 7 Tesla external field. The low value of the ε´ at low temperatures is mainly due to that at relatively low temperatures the electric dipoles cannot orient themselves with the direction of the applied electric field, so their contribution is weak to the polarization[41]. After 150 K, the ε´ increase, which means that after a temperature rise, most of the electric dipoles get enough exciting energy to change the external electric field so the ε´ increases rapidly and reached its maximum position for all frequencies? For the other frequencies 50 kHz, 100 kHz, 500 kHz, and 1 MHz, ε´ is gradually varied from 200 K. In higher temperatures above 200 K, ε´ increases rapidly. It was seen that the dispersion of ε´ at low-frequency 10 kHz is greater than that of higher frequencies. The decrease in dielectric constant value with lower to higher frequency is a universal phenomenon in the manganites system. The higher value of dielectric permittivity at high temperatures was due to the space charge polarization which also affects the charge carriers in the sample. It is interesting to note that the dielectric constant increases when we apply the external magnetic field. From the figure, it is clear that for higher frequency data, the value of dielectric constant at 300 K is 400 at 0 Tesla field wherever after applying 7 Tesla external magnetic field the value is 1700. This increment is also observed for the other frequencies too. The dielectric permittivity increases after applying the field, which improves the dielectric.

**Figure 2(b)** shows the tangential loss (tanδ) of the sample with and without magnetic field in the temperature range of 10 K- 300 K for different frequencies. The behaviour of tanδ is a little bit similar to the dielectric constant. For lower frequencies, tanδ is constant up to 80 K, and after that tanδ increases with an increase in temperature, but for higher frequencies, the value of the tanδ is very low. The increment in the value of tanδ for high temperature indicates that the charge carriers start to activate after increasing temperature, which increases the tanδ value in the sample. When temperature increases the loss in higher frequencies is lower in comparison to lower frequencies. It is clear from **figure 2(a) and (b)** that the dielectric behavior of polycrystalline GPMO at various frequencies with the temperature follows the relaxation mechanisms. For higher frequencies, the low value of ε′ and tanδ suggests that the dipoles orient slowly in the direction of the applied electric field. As the frequency increases the charges cannot follow the field direction and their contribution to the dielectric permittivity has been ended.

### 3.2.2 Magnetocapacitance and magneto loss study:

To check the effect of the magnetic field on the dielectric relaxation mechanism, we studied the Magnetocapacitance and Magnetoloss in the sample for 1 kHz, 5 kHz, 10 kHz, 50 kHz, and 100 kHz frequencies. Magnetocapacitance in the sample can be calculated to determine the power of magnetoelectric coupling in a sample. The Magnetocapacitance (MC) and Magnetoloss (ML) of the sample can be defined as;

$$Magnetocapacitance = \frac{\varepsilon'(H) - \varepsilon'(0)}{\varepsilon'(0)} \times 100\%, \quad Magnetoloss = \frac{tan\delta(H) - tan\delta(0)}{tan\delta(0)} \times 100\%$$

In the above equation, ε′(H), ε′(0), tanδ(H), and tanδ(0) are the dielectric constants, and tangential loss of the sample under the application of magnetic field (7 Tesla) and absence of magnetic field, respectively.

**Figure 2(c)** shows the magneto capacitance and magneto loss data of the polycrystalline GPMO. In Magnetocapacitance data, it is seen that for 10 kHz frequency, there are two dielectric regions, from 100 K-150 K temperature the value increases, and after that, from 150 K- 175 K, there is a plateau region. And after this, there is a sharp increase in data for 175 K- 250 K, which shifts towards higher temperature for the increasing frequencies. When we check the data, for higher frequencies, the Magnetocapacitance data increases and shifts towards higher temperatures. These phenomena suggest that polarizability increases with temperature as well as frequencies. The Magnetocapacitance of the sample is positive for all the frequencies.

Magneto loss data states the loss in capacitance. From the data, it has been confirmed that the nature of tangential loss is the same as the capacitance. For increasing frequency, the data shift towards the higher temperature. Magnetic field-induced strain is used to explain the origin of MC and ML. Material strain is caused by the applied magnetic field. Stress and an electric field are produced as a result of this induced strain. As a result, the dielectric properties of the sample are modified.

### 3.2.3 Electric Modulus study:

To explain the space charge relaxation phenomena in the materials, Macedo et al. explained the concept of complex electric modulus[42]. The complex electric modulus ($M^* = M' + jM''$) can be expressed as the change in complex permittivity form by using the relation, $M^* = \frac{1}{\varepsilon^*} = \frac{\varepsilon'}{\varepsilon'^2 + \varepsilon''^2} + j\frac{\varepsilon''}{\varepsilon'^2 + \varepsilon''^2} = M' + jM''$. A complex electric modulus study is the easiest and most appropriate way to analyze the transport phenomena in the materials. In addition, it gives some critical information, such as the hopping rate of ions or charge carriers and the relaxation time for conductivity. **Figure 3(a), and (b),** show the real (M´) and imaginary (M´´) parts of the electric modulus with and without field, respectively. From the figure, it has been seen that the real and imaginary part of the modulus decreases with an increase in temperature for all the frequencies, which confirms that the trend of electric modulus is almost the same for 0 Tesla and 7 Tesla data. Eroglu et al. proposed that in a wide range of temperatures, when the real electric modulus of the device decreased, the imaginary part of the electric modulus increased. In our study, the real and imaginary part of the electric modulus decreases from its maximum values, when the temperature increases. These results confirm that synthesized polycrystalline GPMO has a hopping mechanism and for the electrical conduction mechanism the surface charge polarization exists. Also, the modulus peaks are shifting towards a higher temperature for low frequencies to higher frequencies, which indicate the thermally activated behavior in the polycrystalline GPMO. Other types of manganites also showed these types of behavior. Multiple carriers in localized conduction are responsible for this polarization mechanism. This confirms the relaxation processes in the GPMO sample. The variation of the modulus indicates the broad and asymmetric peaks in the materials, which indicates the Non-Debye type relaxation in the synthesized material. The shifting of peaks to the higher temperature side for all the frequencies is due to the correlated motions of mobile ions[43]. It can be said that the real part of the modulus goes up with a higher frequency with increasing temperature. The higher value of M´ around 120 K for all the frequencies suggest that the particle size of the material is small. The variation of M´´ with the temperature at different frequencies shows that dispersion shifts towards the higher value of frequency. It is confirmed from the results that, As a result of thermal activation, the hopping mechanism of charge carriers dominates the dielectric relaxation process. It has been seen that relaxation peaks shift from lower frequency to higher frequency when the temperature is increased. This type of movement of the peaks can be linked to the thermal energy of charge carriers.

### 3.2.4 Conductivity and Activation energy Analysis:

To confirm the relaxation mechanism of the GPMO sample, we have done the temperature-dependent ac conductivity at different frequencies for 0 Tesla and 7 Tesla external magnetic fields (**figure 4(a) and (b)**). To determine the conduction mechanism in the materials, dielectric relaxation is the important parameter. As we know that ac conductivity is dependent upon the dielectric constant, the ac conductivity behavior is the same as that of the dielectric constant. It is seen in the figure that for lower frequencies up to 30 kHz, ac conductivity is constant. In higher frequencies between 55 kHz to 1 MHz, ac conductivity gradually increases with an increase in temperature. A slow electric field reversal in lower frequency at a lower temperature is related to the low conductivity in the sample due to the accumulation of ions. It has been observed in the figure that for a higher frequency, the ac conductivity increases due to the small polaron mechanism. This result confirms that the dielectric relaxation peaks and conduction phenomena in the sample belong to the same class, which demonstrates that relaxation phenomena are related to the conductivity of the grain. From the figure, it is noted that the conductivity increases after applying an external magnetic field.

The inset of **Figures 4(a) and (b),** show the ln conductivity with the reverse of temperature, as in the Arrhenius plots. In the relaxation process, the activation energy can be determined by the Arrhenius plot. The data shows the semiconducting-like behaviour, as the calculated activation energy lies in the small range as that of semiconductors. From lower frequency to higher frequencies the ac conductivity increases linearly with an increase in temperature, confirming that it is the thermally activated mechanism for ac conductivity [44]. The increase in ac conductivity–as temperature increases means the trapped charges are unconventional by which electrons exchange is increased. By increasing the temperature, the dispersion in conductivity decreases. The conductivity values seen at high temperatures are slightly frequency-independent. The slopes of ln ac conductivity versus 1000/T were further used to calculate the activation energy by linear fitting for frequencies. The calculated activation energy lies between 0.080 eV to 0.095 eV for 500 Hz frequency to 1 MHz frequency, respectively for the 0 Tesla field data. Also, the activation energy for the externally applied field varies from 0.080 eV to 0.090 eV for 500 Hz frequency to 1 MHz frequency. The activation energy has less value of energy than 1 eV, which confirms the electronic conduction mechanism in the present sample.

**3.3 Temperature-dependent structural study using Synchrotron X-Ray Diffraction:**

Structural correlation to magneto-dielectric properties is important to investigate. It is obvious from the magneto-dielectric study that there is no ferroelectric transition at lower temperature without magnetic field which can be considered as no further structural transition at lower temperature. We have performed the temperature-dependent high-resolution Synchrotron angle dispersive x-ray diffraction measurements from 300 K to 50 K temperatures for understanding the structural changes in polycrystalline GPMO. In **figure 5**, the diffraction patterns of polycrystalline GPMO are shown at 300 K to 50 K using synchrotron angle dispersive x-ray diffraction (AD-XRD). X-ray diffraction data were collected on a prepared sintered polycrystalline GPMO sample from 300 K to 50 K within the 2θ region of 10˚-35˚ for the wavelength λ= 0.81541 Å, shown in **figure 6(a)**. The spectra of GPMO were indexed with an orthorhombic structure having a *Pnma* space group. No intermediate or secondary phase is detected in the peak match, indicating a single-phase formation. The (212) peak is split into two peaks at 50 K XRD data. Moreover, the peak (331) is shifting on a higher angle side from 300 K to 50 K shown in the inset of **figure 5**. A Rietveld refinement of the AD-XRD data was performed using the computational software JANA2006 in order to understand any subtle changes in the structure and quantify the occupancy of lattice sites, shown in **figure 6**. We used the orthorhombic *Pnma* space group for refinement at all temperatures. It is evident that peaks in **figure 5**, a little bit shifts toward lower 2theta values with increasing temperature, which is expected due to thermal expansion. Thermal expansion or ion ordering in the shell configuration of rare earth ions with different radii ($Gd^{3+}$: 1.107Å and $Pr^{3+}$: 1.117Å ) is expected to cause the phenomenon[45]. With reasonably good fit parameters ($R_p$, $wR_p$, and GOF), there is a good agreement between the observed and calculated patterns. A detailed overview of cell parameters and R-factors can be found in table 2. Polyhedral views of polycrystalline GPMO for 300 K, 200 K, 100 K, and 50 K are shown in **Figures 7(a) and 7(b)** showing the top view of the octahedral tilting in the crystal structure of GPMO drawn with VESTA software. The values of the structure's atomic positions along with bond length and bond angle are shown in **table 2**. It is confirmed from the structure analysis that there is no change in the phase formation in the polycrystalline GPMO from room temperature to low temperature.

Based on the lattice parameters of the sample at different temperatures, orthorhombic strain can be calculated for the GPMO sample. The calculated orthorhombic strain of the *Pnma* space group is 0.04133(1), 0.04142(8), 0.04149(6), and 0.04152(4) for 300 K, 200 K, 100 K, and 50 K, respectively. The values confirmed that from going 300 K to 50 K strain is increased. A-site doping increases the average ionic radius of the A-site ion, which results in a decrease in tilt and an increase in Mn-O-Mn bond angle. The calculated value of octahedral tilting combined with the static Jahn teller (JT) distortion produced by the superexchange interaction of $Mn^{3+}$-O-$Mn^{3+}$, is present in **table 2**. In this way, GPMO possessed an orthorhombic crystal structure having a *Pnma* space group for 300 K as well as for 50 K.

### 3.4 Magnetic study:

The temperature-dependent *dc* magnetization measurement has been collected for the magnetic phase transition in polycrystalline GPMO at 100 Oe magnetic fields. Generally, When the materials are cooled down from room temperature to low temperature in rare-earth manganites, the $Mn^{3+}$ sublattice ordering is observed at Neel temperature $T_N$, which decrease monotonically from higher ionic radii to lower ionic radii compound such as 145 K in $LaMnO_3$ to 45 K in $HoMnO_3$, although the $R^{3+}$ sublattice orders at a much lower temperature, normally below 20 K. So the Mn ordering temperature should lie in this temperature range as we substituted $Pr^{3+}$ at the site of $Gd^{3+}$. In **figure 8(a)**, the temperature variation of the moment in the sample under an external magnetic field of 100 Oe is shown at the zero-field cooled and field-cooled states. During the experiment, the sample was cooled to 5 K and a magnetic field of 100 Oe was applied. A warming run was conducted after zero-field cooling data were measured up to 300 K, while applied magnetic fields cooled the system again and the field-cooled (FC) magnetization data were recorded. In the figure, the ZFC and FC curves are seen to diverge at around 65 K, following which magnetic ordering begins to take place in the material.

In the temperature range of 60 K-35 K, the ZFC curve is showing purely antiferromagnetic property of the material due to Mn spin ordering. From the first derivative of ZFC magnetization (dM/dT curve in Fig 9 (c)), the magnetic transition can be observed at $T_N$ = 59 K where the $Mn^{3+}$ spins are started to getting ordered antiferromagnetically. Second magnetic transition can be observed at $T_c$ = 43 K, which is critical magnetic transition temperature where the material is started to approach the negative magnetization. The negative magnetization occurs because the average effective spin of Gd ions is greater than 7/4 whereas the average effective spin of the Mn ions is less than 7/4 and the spin alignment of Gd ions aligns antiparallel to the spin alignment of the Mn ions in the direction of weak external magnetic field (H=100 Oe). Below 30 K, the spins of $Pr^{3+}$ and $Gd^{3+}$ magnetic lattice is started contributing and ordered ferromagnetically for which the magnetic moment is exceeds the magnetic moment of Mn ions and shows the spin reversal phenomena with negative magnetization. FC curve is as expected giving larger magnetic moment with antiferromagnetic behavior of the material in the temperature range of 65 K- 35K and then it is crossing the ZFC curve at 35 K where the magnetization reversal phenomena occurs. A closer look at the parent samples reveals that neither $GdMnO_3$ nor $PrMnO_3$ exhibit negative magnetization. The solid solution of these two compounds exhibits this property. The compound may undergo weak anisotropy around the Gd site after $Pr^{3+}$ is replaced at the $Gd^{3+}$ site, which is associated with low-temperature negative magnetization. A negative magnetization occurs when the temperature goes from high to low and the Gd and Pr spins polarize opposite the exchange field of the Mn moment[46]. In the case of low temperatures, sublattice magnetization is antiparallel to the field. According to Curie-Weiss law, antiferromagnetic/ferromagnetic materials have different magnetic susceptibility at absolute temperatures (T) depending on:

$$\frac{1}{\chi} = \frac{T - T_\theta}{C}$$

where C and $T_\theta$ are Curie Constant and Curie-Weiss temperature, respectively. The linear fit of $1/\chi$ vs. T data in the 100 K-300 K temperature range (paramagnetic region) is shown in Fig. 9(b). Above a certain temperature nearly 70 K, it remains nearly constant, however below this temperature, it starts increasing rapidly, indicating a change in magnetic behavior. From the Curie-Weiss fitting analysis, a negative Curie-Weiss temperature $T_\theta$ = -11.612 K value indicates that magnetic ordering should be antiferromagnetic. The theoretical paramagnetic effective magnetic moment $\mu_{effective}$ for GPMO can be calculated as $\sqrt{(0.55\mu_{Gd^{3+}}^2) + (0.45\mu_{Pr^{3+}}^2) + (0.55\mu_{Mn^{3+}}^2) + (0.45\mu_{Mn^{3+}}^2)}$ where $\langle\mu_{Gd}^{3+}\rangle = 7.94\mu_B$, $\langle\mu_{Pr}^{3+}\rangle = 3.58\mu_B$ and $\langle\mu_{Mn}^{3+}\rangle = 4.89\mu_B$. The experimentally calculated value for GPMO is 7.2304(5) $\mu_B$ which is nearly equal to the theoretical value of GPMO 8.022(6) $\mu_B$. Below ~60 K, it is attaining the Griffiths phase which is continuing down to 25 K, after that the magnetization spin reversal phenomena was observed in the material which is really a matter of interest. At lower temperature, around 5 K-10 K, there is sharp increase in the moment which may be due to the ferromagnetic clustering of the Gd+Pr magnetic spin.

The field-dependent magnetization measurements have been performed at 300 K and 5 K, on polycrystalline GPMO. The measurement has been carried out after the cooling of the sample under zero fields. The result is shown in **figure 9(a)**. It is shown in the figure that at room temperature M-H curve shows the paramagnetic behavior the same as the parent sample $GdMnO_3$. For low-temperature Pr doped $GdMnO_3$ shows the step like meta-magnetic transition. It is also seen that the M-H loop is non- saturating in nature and the magnetic transition is very broad. The large coercivity of 6592.77 Oe from the hysteresis loop ($H_c$) was observed at 5 K. As discussed the magnetic behavior in case of temperature dependent magnetization study, broad hysteresis at 5 K representing the combination of ferromagnetic domains of Gd+Pr magnetic lattice with canted anti-ferromagnetic domains of Mn lattice. Polycrystalline GPMO can be a promising candidate for the magnetic switching and memory device applications. A detailed magnetic behavior of GPMO for understanding the negative magnetization and its behavior on higher field is obviously a matter of research. A future work will be focused on the series of Pr doped GdMnO$_3$ samples for understanding the structural correlation to magnetic properties of the sample at low temperatures to develop the magnetic switching properties in manganites.

4. Conclusion:

The solid-state synthesis method was successfully used to synthesize polycrystalline GPMO at higher temperatures. Rietveld refinement of synchrotron AD-XRD confirms the purity of the sample having an orthorhombic crystal structure with a *Pnma* space group. Dielectric permittivity data of the sample with and without an external magnetic field shows the temperature-dependent behavior of the sample. The permittivity data are influenced by dielectric relaxation phenomena. The dielectric behavior of the sample was enhanced after applying the 7 Tesla external magnetic field. The *ac* conductivity of the sample was also enhanced after applying the external magnetic field and the calculated value of activation energy conforms to the increasing trend of the activation energy from lower frequencies to higher frequencies. The Dc magnetization study confirms the metamagnetic behavior in the polycrystalline GPMO.

**Acknowledgment:**

The authors would like to thank Prof. Debanand Sa for his valuable discussion. The authors want to acknowledge Dr. R. J. Chaudhary UGC- DAE- CSR Indore, India for the magnetic measurement. PP and HA are thankful to and Dr. Devendra Pandey and Dr. Parul for scientific discussion on magnetic measurement. MAS want to acknowledge IOE, Banaras Hindu University, India for financial support.

**References:**

[1]     Cai X, Shi L, Zhou S, Zhao J, Guo Y and Wang C 2014 Size-dependent structure and magnetic properties of DyMnO3 nanoparticles *Journal of Applied Physics* **116**

[2]     Xia W, Pei Z, Leng K and Zhu X 2020 Research Progress in Rare Earth-Doped Perovskite Manganite Oxide Nanostructures *Nanoscale Research Letters* **15**

[3]     Ramesh R and Spaldin N A 2007 Multiferroics: Progress and prospects in thin films *Nature Materials* **6** 21–9

[4]     Wang K F, Liu J M and Ren Z F 2009 Multiferroicity: The coupling between magnetic and polarization orders *Advances in Physics* **58** 321–448

[5]     Goto T, Kimura T, Lawes G, Ramirez A P and Tokura Y 2004 Ferroelectricity and giant magnetocapacitance in perovskite rare-earth manganites *Physical Review Letters* **92**

[6]     Tomioka Y and Tokura Y 2004 Global phase diagram of perovskite manganites in the plane of quenched disorder versus one-electron bandwidth *Physical Review B - Condensed Matter and Materials Physics* **70**

[7]     Doshi R R, Solanki P S, Khachar U, Kuberkar D G, Krishna P S R, Banerjee A and Chaddah P 2011 First order paramagneticferromagnetic phase transition in Tb3 doped La0.5Ca0.5MnO3 manganite *Physica B: Condensed Matter* **406** 4031–4

[8]     Sharma H, Kumar D, Tulapurkar A and Tomy C V. 2019 Effect of B-site bismuth doping on magnetic and transport properties of La0.5Ca0.5Mn1−xBixO3 thin films *Journal of Materials Science* **54** 130–8

[9]     Wang Z M, Ni G, Sang H and Du Y W 2001 The effect of average A-site cation radius on TC in perovskite manganites *Journal of Magnetism and Magnetic Materials* **234** 213–7

[10]    Sapana S, Dhruv D, Joshi Z, Gadani K, Rathod K N, Boricha H, Shrimali V G, Trivedi R K, Joshi A D, Pandya D D, Solanki P S and Shah N A 2017 Studies on structural and electrical properties of nanostructured RMnO3 (R = Gd & Ho) *AIP Conference Proceedings* vol 1837

[11]    Kimura T, Ishihara S, Shintani H, Arima T, Takahashi T, Ishizaka K and Tokura Y 2003 Distorted perovskite with eg1 configuration as a frustrated spin system *Physical Review B - Condensed Matter and Materials Physics* **68**

[12]    Dong S, Yu R, Yunoki S, Liu J M and Dagotto E 2008 Origin of multiferroic spiral spin order in the R MnO3 perovskites *Physical Review B - Condensed Matter and Materials Physics* **78**

[13]    Kim B H and Min B I 2009 Nearest and next-nearest superexchange interactions in orthorhombic perovskite manganites R MnO3 (R=rare earth) *Physical Review B -*

*Condensed Matter and Materials Physics* **80**

[14]   Liu J M and Dong S 2015 Ferrielectricity in DyMn2O5: A golden touchstone for multiferroicity of RMn2O5 family *Journal of Advanced Dielectrics* **5** 1–18

[15]   Huang Z, Cao Y, Sun Y, Xue Y and Chu C 1997 Coupling between the ferroelectric and antiferromagnetic orders *Physical Review B - Condensed Matter and Materials Physics* **56** 2623–6

[16]   Kimura T, Goto T, Shintani H, Ishizaka K, Arima T and Tokura Y 2003 Magnetic control of ferroelectric polarization *Nature* **426** 55–8

[17]   Gadani K, Dhruv D, Joshi Z, Boricha H, Rathod K N, Keshvani M J, Shah N A and Solanki P S 2016 Transport properties and electroresistance of a manganite based heterostructure: Role of the manganite-manganite interface *Physical Chemistry Chemical Physics* **18** 17740–9

[18]   Arayedh B, Kallel S, Kallel N and Peña O 2014 Influence of non-magnetic and magnetic ions on the MagnetoCaloric properties of La0.7Sr0.3Mn0.9M 0.1O3 doped in the Mn sites by M=Cr, Sn, Ti *Journal of Magnetism and Magnetic Materials* **361** 68–73

[19]   Zhao R, Jin K, Xu Z, Guo H, Wang L, Ge C, Lu H and Yang G 2013 The oxygen vacancy effect on the magnetic property of the LaMnO 3-δ thin films *Applied Physics Letters* **102**

[20]   Hemberger J, Brando M, Wehn R, Ivanov V Y, Mukhin A A, Balbashov A M and Loidl A 2004 Magnetic properties and specific heat of RMnO3 (R=Pr, Nd) *Physical Review B - Condensed Matter and Materials Physics* **69** 1–5

[21]   Markovich V, Fita I, Mogilyansky D, Wisniewski A, Puzniak R, Titelman L, Vradman L, Herskowitz M and Gorodetsky G 2007 Magnetic properties of nanocrystalline La1-xMnO 3+δ manganites: Size effects *Journal of Physics Condensed Matter* **19**

[22]   Iqbal M, Khan M N and Khan A A 2018 Structural, magnetic, magnetocaloric and critical behavior studies in the vicinity of the paramagnetic to ferromagnetic phase transition temperature in LaMnO3+δ compound *Journal of Magnetism and Magnetic Materials* **465**

670–7

[23]   Zhang G Q, Luo S J, Dong S, Gao Y J, Wang K F and Liu J M 2011 Enhanced ferroelectricity in orthorhombic manganites Gd 1-xHoxMnO3 *Journal of Applied Physics* **109** 107–10

[24]   Sakka A, M'nassri R, Chniba-Boudjada N, Ommezzine M and Cheikhrouhou A 2016 Effect of trivalent rare earth doping on magnetic and magnetocaloric properties of Pr0.5(Ce,Eu,Y)0.1Sr0.4MnO3 manganites *Applied Physics A: Materials Science and Processing* **122** 1–12

[25]   Peña O, Bahout M, Ghanimi K, Duran P, Gutierrez D and Moure C 2002 Spin reversal and ferrimagnetism in (Gd, Ca)MnO3 *Journal of Materials Chemistry* **12** 2480–5

[26]   Zhang R R, Kuang G L, Yin L H and Sun Y P 2012 Magnetization reversal in the Cr-doped manganite Bi 0.3Ca 0.7Mn 0.75Cr 0.25O 3 *Journal of Alloys and Compounds* **519** 92–6

[27]   Sharma N, Srivastava B K, Krishnamurthy A and Nigam A K 2010 Magnetic behaviour of the orthochromite La0.5Gd 0.5CrO3 *Solid State Sciences* **12** 1464–8

[28]   Yoshii K, Nakamura A, Ishii Y and Morii Y 2001 Magnetic properties of La1-xPRxCRO3 *Journal of Solid State Chemistry* **162** 84–9

[29]   Hemberger J, Lobina S, Krug Von Nidda H A, Tristan N, Ivanov V Y, Mukhin A A, Balbashov A M and Loidl A 2004 Complex interplay of 3d and 4f magnetism in La1-xGd xMnO3 *Physical Review B - Condensed Matter and Materials Physics* **70**

[30]   Panwar N, Joby J P, Kumar S, Coondoo I, Vasundhara M, Kumar N, Palai R, Singhal R and Katiyar R S 2018 Observation of magnetization reversal behavior in Sm0.9Gd0.1Cr0.85Mn0.15O3 orthochromites *AIP Advances* **8**

[31]   Pant P, Agarwal H, Bharadwaj S and Shaz M A 2022 Effects of Cr and Fe Substitution at Mn-Sites of Gdmn1-Xtxo3 (X=0, 0.10) on its Structural and Complex Dielectric Properties *SSRN Electronic Journal* **290**

[32]   Karmakar S, Tyagi H, Mohapatra D P and Behera D 2021 Dielectric relaxation behavior and overlapping large polaron tunneling conduction mechanism in NiO–PbO μ-cauliflower composites *Journal of Alloys and Compounds* **851**

[33]   Kimura T, Lawes G, Goto T, Tokura Y and Ramirez A P 2005 Magnetoelectric phase diagrams of orthorhombic RMnO3 (R=Gd, Tb, and Dy) *Physical Review B - Condensed Matter and Materials Physics* **71**

[34]   Goto T, Yamasaki Y, Watanabe H, Kimura T and Tokura Y 2005 Anticorrelation between ferromagnetism and ferroelectricity in perovskite manganites *Physical Review B - Condensed Matter and Materials Physics* **72**

[35]   Schrettle F, Lunkenheimer P, Hemberger J, Ivanov V Y, Mukhin A A, Balbashov A M and Loidl A 2009 Relaxations as key to the magnetocapacitive effects in the perovskite manganites *Physical Review Letters* **102** 2–5

[36]   Agarwal H, Alonso J A, Muñoz Á, Choudhary R J, Srivastava O N and Shaz M A 2020 Evolution from spiral to canted antiferromagnetic spin-ordered magnetic phase transition in Tb0.6Pr0.4MnO3

[37]   Pant P, Agarwal H, Bharadwaj S, Srivastava O N and Shaz M A 2021 Relaxor ferroelectric phase transition and ac conduction in polycrystalline Gd0.55Ca0.45MnO3 at low temperature *Materials Chemistry and Physics* **267**

[38]   Vegard L 1921 Die Konstitution der Mischkristalle und die Raumfüllung der Atome *Zeitschrift für Physik* **5** 17–26

[39]   Shannon R D 1976 Revised effective ionic radii and systematic studies of interatomic distances in halides and chalcogenides *Acta Crystallographica Section A* **32** 751–67

[40]   Knížek K, Jirák Z, Pollert E, Zounová F and Vratislav S 1992 Structure and magnetic properties of Pr1-xSrxMnO3 perovskites *Journal of Solid State Chemistry* **100** 292–300

[41]   Shafqat M B, Ali M, Atiq S, Ramay S M, Shaikh H M and Naseem S 2019 Structural, morphological and dielectric investigation of spinel chromite (XCr2O4, X = Zn, Mn, Cu & Fe) nanoparticles *Journal of Materials Science: Materials in Electronics* **30** 17623–9

[42]    Macedo P B, Moynihan C T and Bose R 1972 Role of Ionic Diffusion in Polarization in Vitreous Ionic Conductors. *Physics and Chemistry of Glasses* **13** 171–9

[43]    Borsa F, Torgeson D R, Martin S W and Patel H K 1992 Relaxation and fluctuations in glassy fast-ion conductors: Wide-frequency-range NMR and conductivity measurements *Physical Review B* **46** 795–800

[44]    Morsi M M, Margha F H and Morsi R M M 2018 Effect of sintering temperature on the developed crystalline phases, optical and electrical properties of 5ZnO-2TiO2- 3P2O5 glass *Journal of Alloys and Compounds* **769** 758–65

[45]    Li W, Liang X, An P, Feng X, Tan W, Qiu G, Yin H and Liu F 2016 Mechanisms on the morphology variation of hematite crystals by Al substitution: The modification of Fe and O reticular densities *Scientific Reports* **6** 1–10

[46]    Kumar A and Yusuf S M 2015 The phenomenon of negative magnetization and its implications *Physics Reports* **556** 1–34

**Figures:**

Figure 1: (a) X-ray diffraction data of polycrystalline GPMO investigated by lab source setup. Inset shows the orthorhombic crystal structure of polycrystalline GPMO. (b) Polyhedral view of crystal structure, (c) top view of the structure (d) bond angle presentation, and (e) bond length presentation of GPMO

Figure 2: (a) (b) Temperature-dependent dielectric constant and tangential loss of the sample with and without magnetic field. (c) Magnetocapacitance and magnetoloss data of polycrystalline GPMO

Figure 3: Temperature- dependent real and imaginary part of electric modulus at 0 Tesla and 7Tesla external magnetic field.

Figure 4: (a) Temperature versus ac conductivity at different frequencies. (b) An Arrhenius plot of temperature dependence conductivity.

Figure 5: Synchrotron Angle dispersive X-ray diffraction data of polycrystalline $Gd_{0.55}Pr_{0.45}MnO_3$ at room temperature to low temperatures.

Figure 6: Rietveld refinement data of polycrystalline Gd0.55Pr0.45MnO3 at 300K, 200K, 100K, and 50K.

Figure 7: (A) Polyhedral crystal structure view of polycrystalline $Gd_{0.55}Pr_{0.45}MnO_3$ at different temperatures, (B) top view of the crystal structure of $Gd_{0.55}Pr_{0.45}MnO_3$ at different temperatures.

Figure 8: (a) dc magnetization curve of ZFC FC data of polycrystalline GPMO, (b) inverse susceptibility curve of Curie Weiss fitting

Figure 9: M-H curve of polycrystalline GPMO at 300 K and 5 K temperatures.

**Tables:**

Table 1: (a) Refined lattice parameter of polycrystalline $Gd_{0.55}Pr_{0.45}MnO_3$ (b) structural parameter of $Gd_{0.55}Pr_{0.45}MnO_3$ sample refined by Rietveld refinement analysis.

Table 2: (a) Refined lattice and (b) atomic parameters of AD-XRD data of polycrystalline GPMO

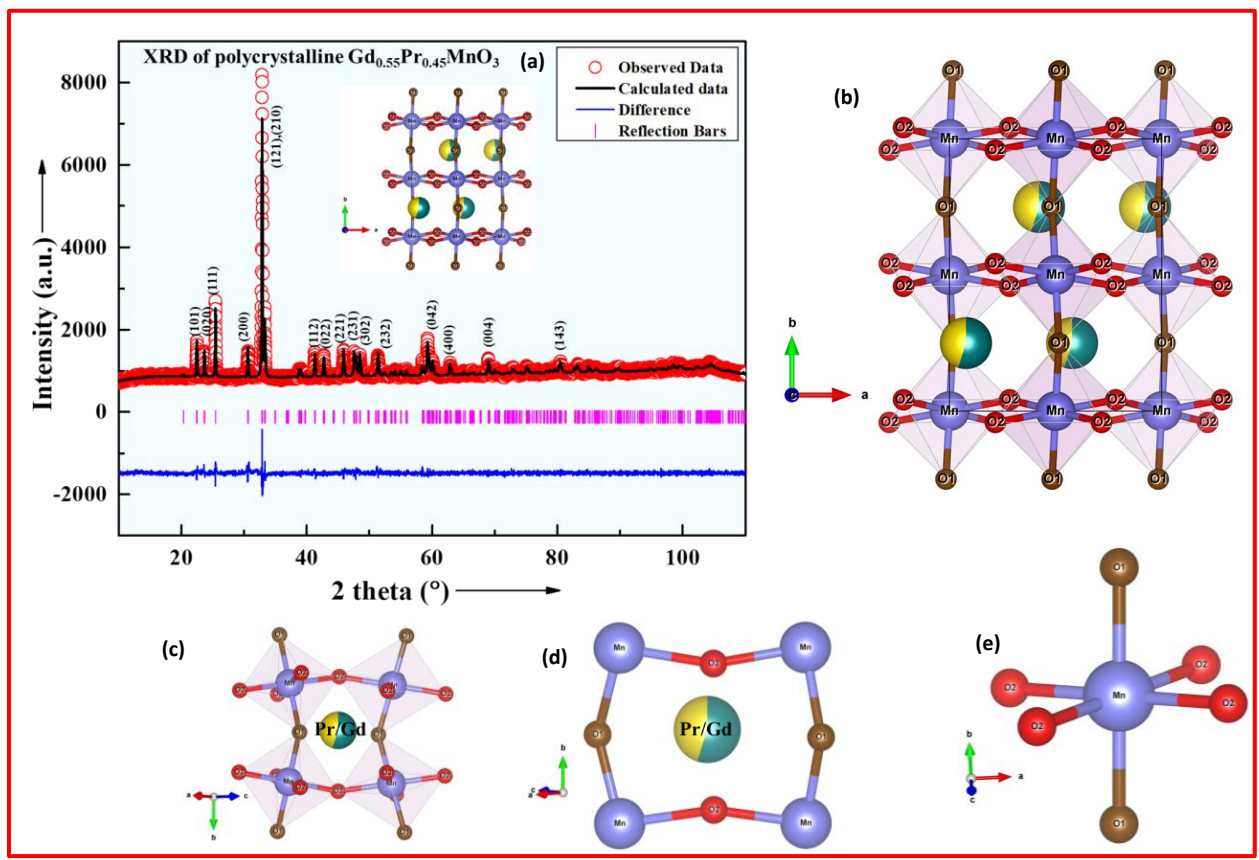

**Figure 1**

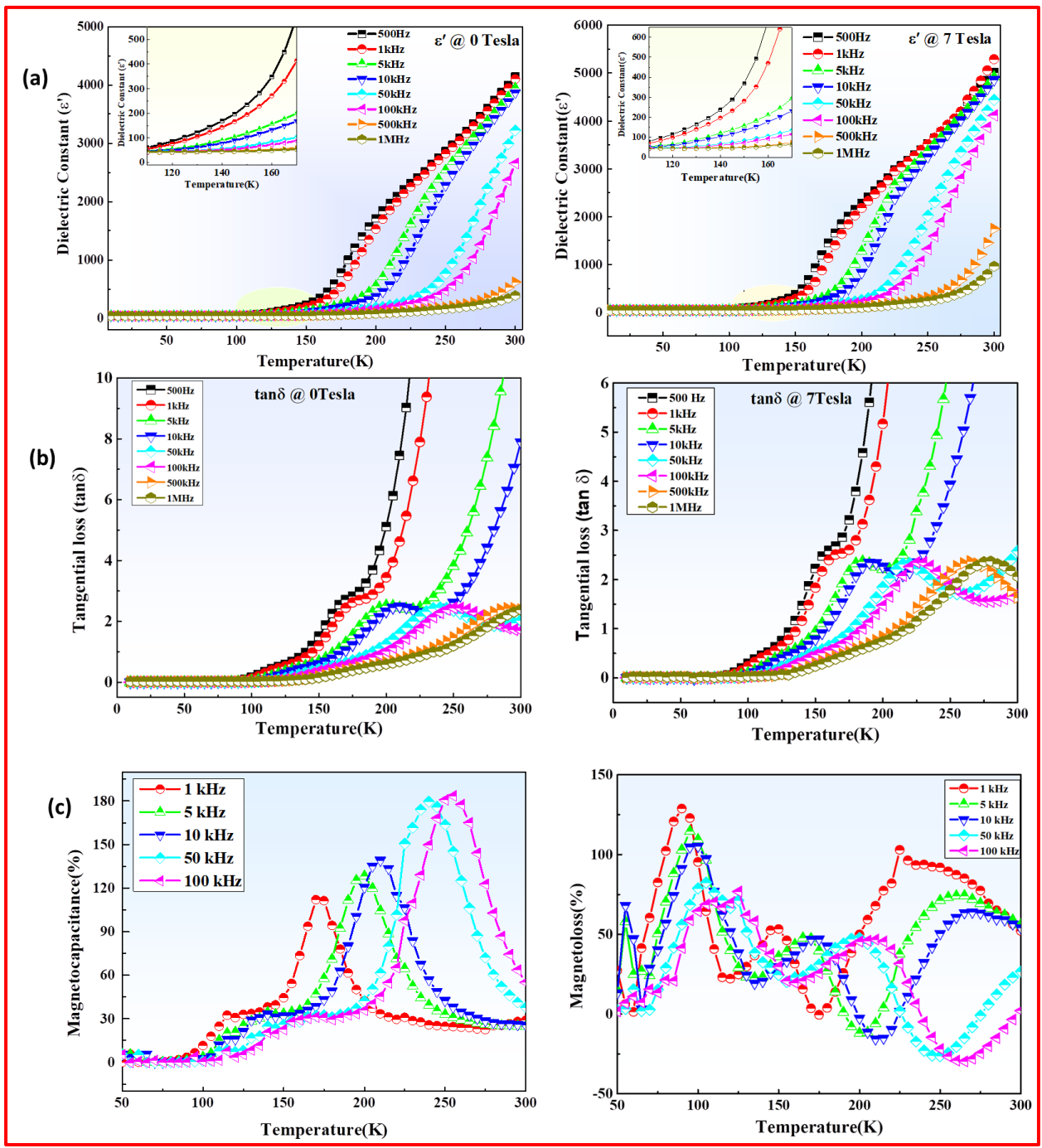

**Figure 2**

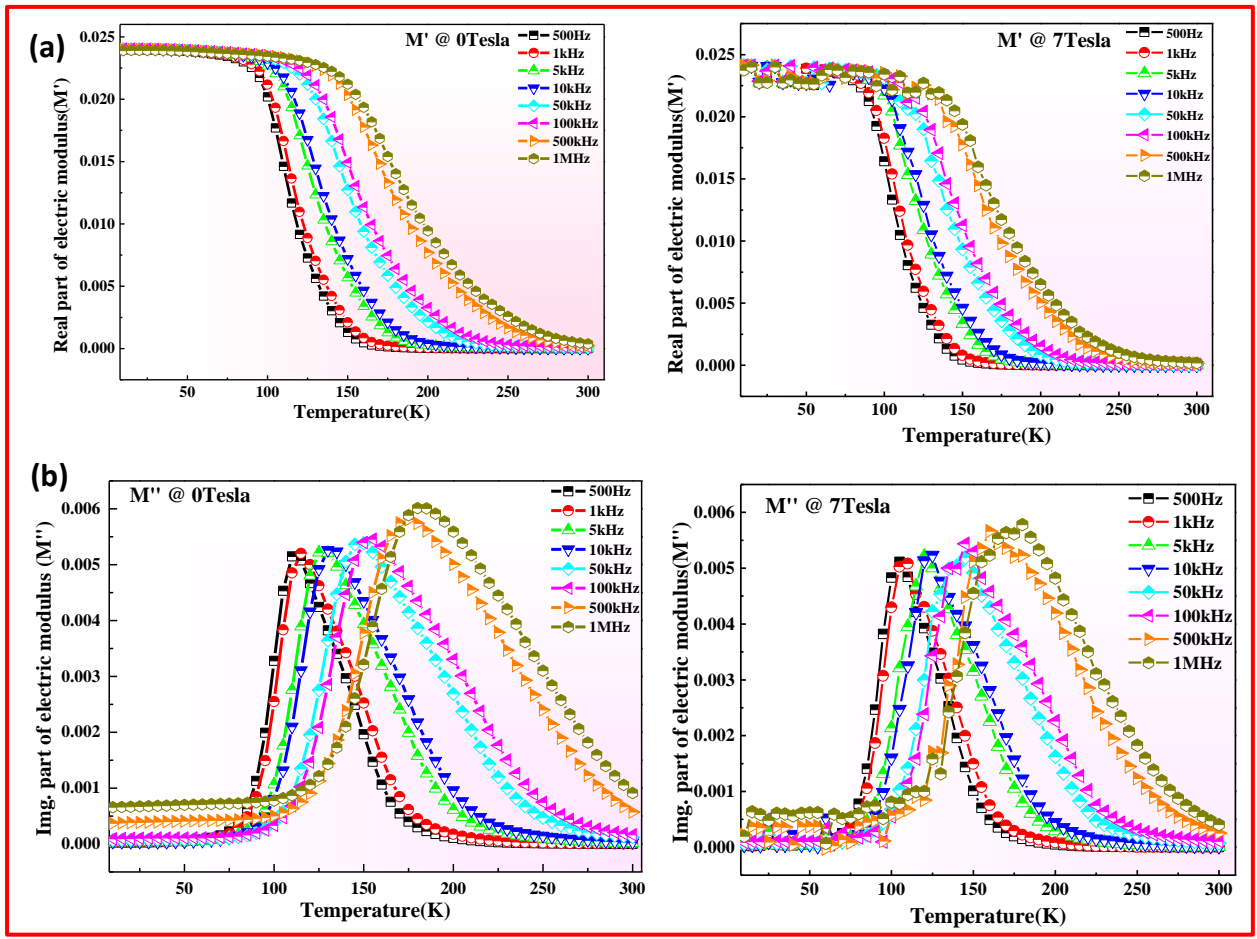

Figure 3

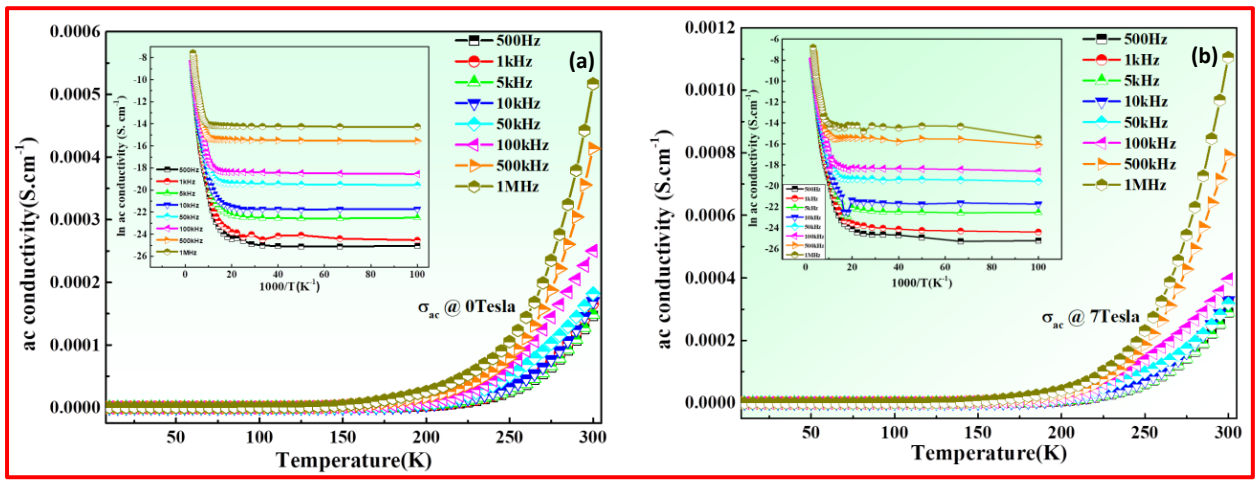

Figure 4

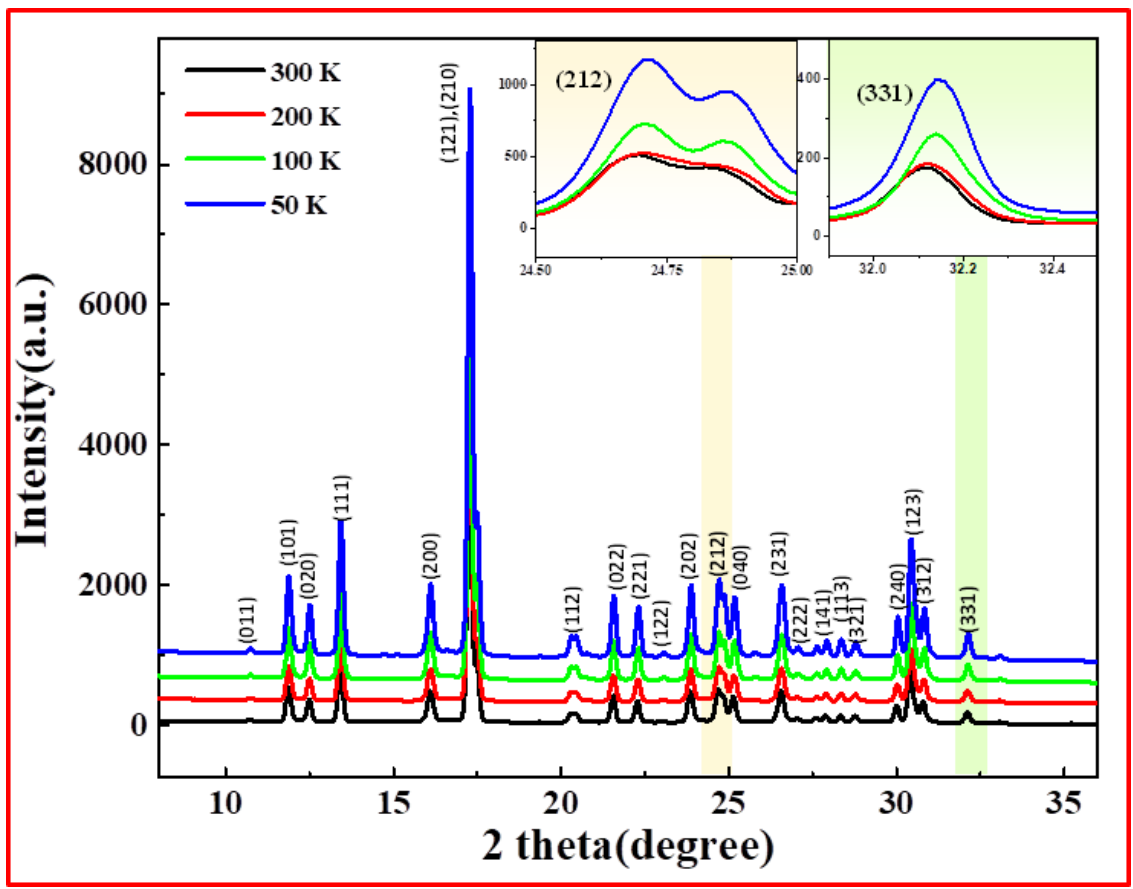

**Figure 5**

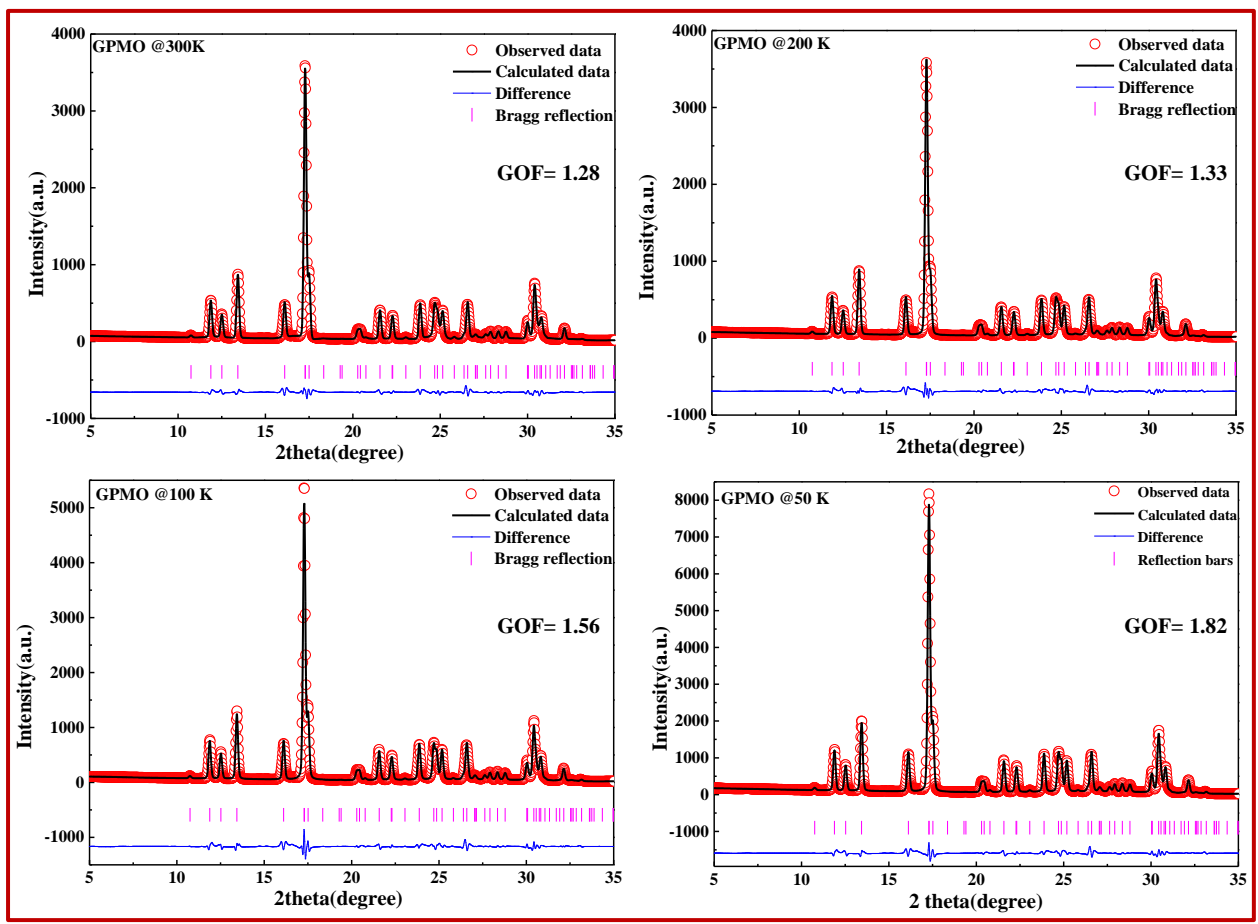

**Figure 6**

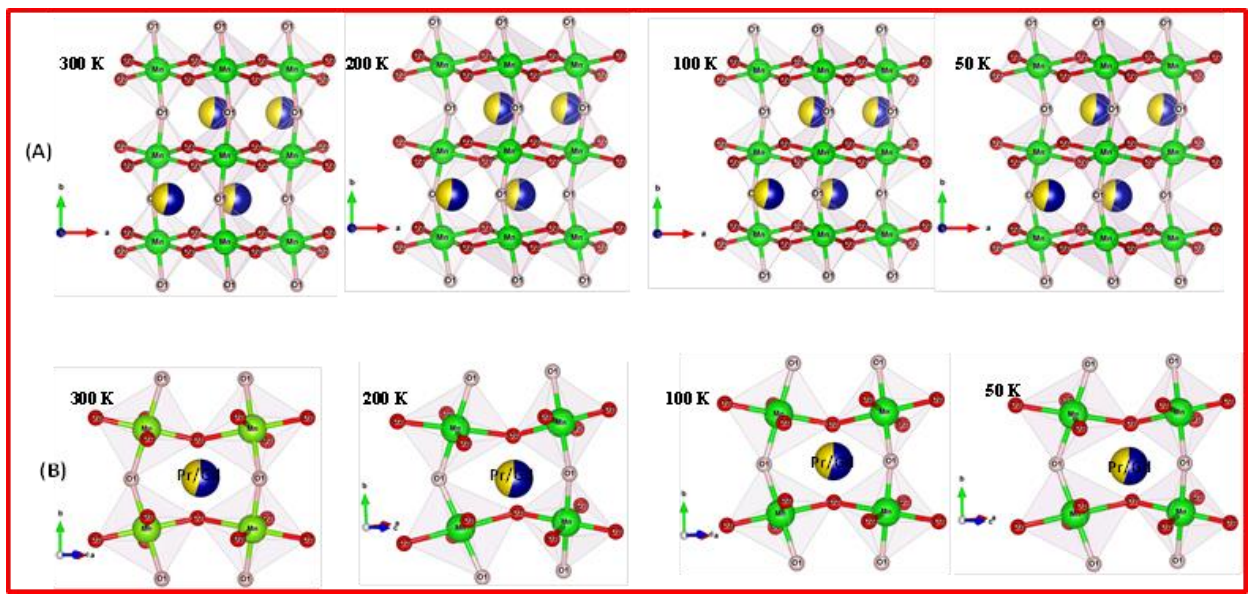

**Figure 7**

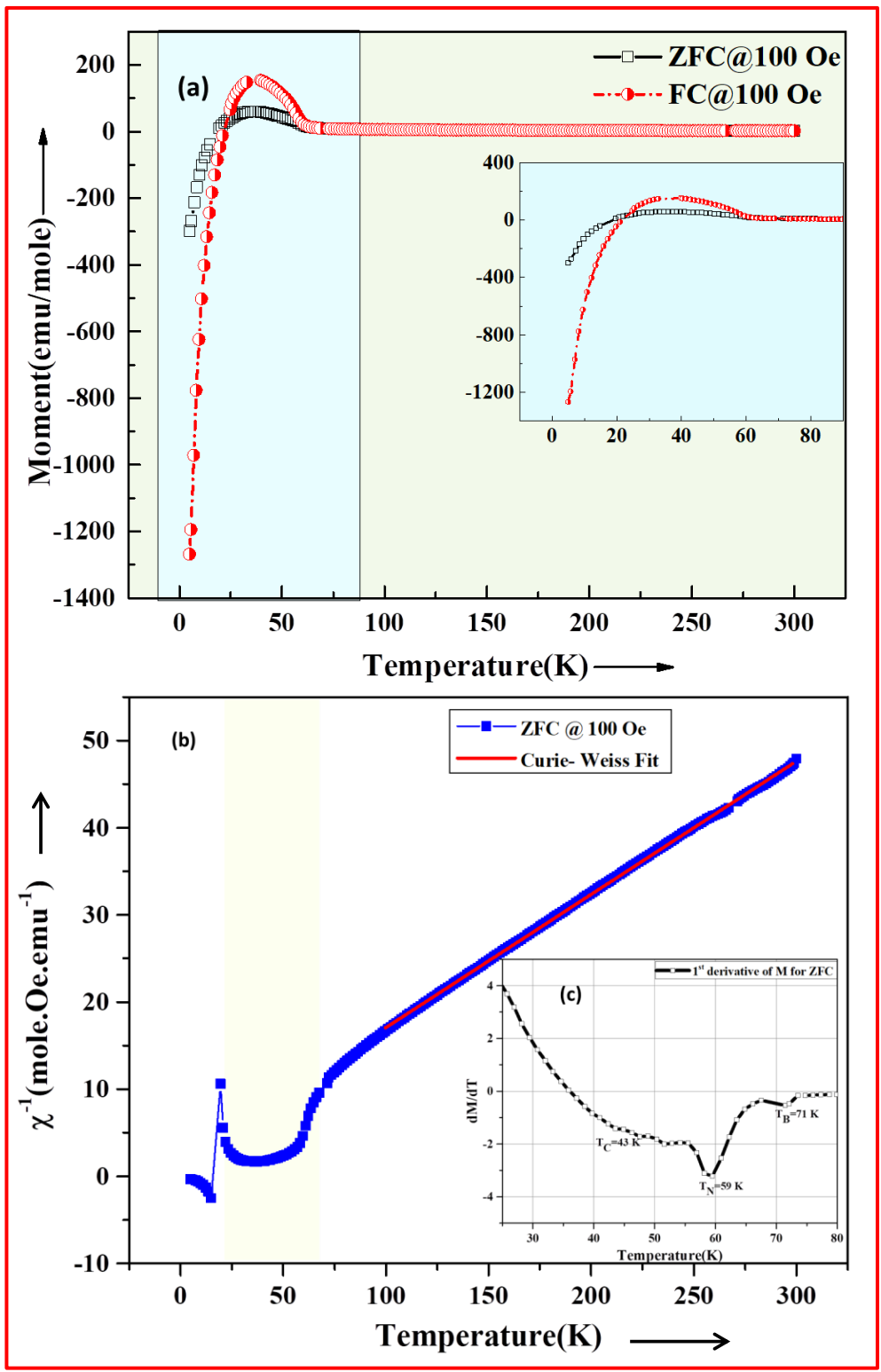

**Figure 8**

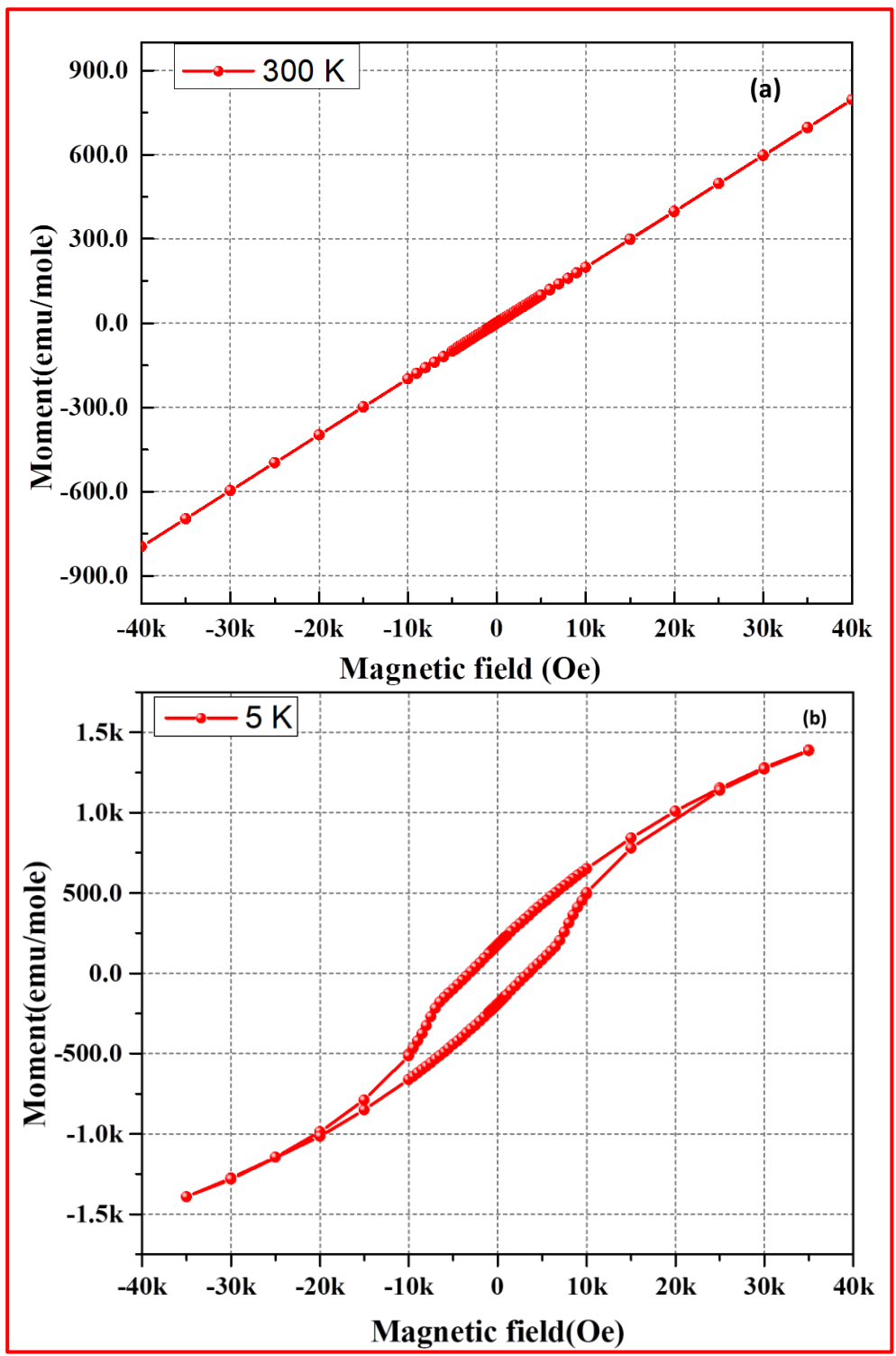

**Figure 9**

| Table 1(a): Refined Parameters and Phase Data of polycrystalline GPMO ||
|---|---|
| **Cell Parameters(Å)** | a = 5.8328(1), b= 7.5041(1), 5.3708(2), <br> α=β=γ= 90° |
| **Space Group** | Pnma |
| **Cell Volume(Å$^3$)** | 235.07944(9) |
| **Profile R-factors** | GOF= 1.10, R$_p$= 2.63%, w$_{Rp}$= 3.54% |
| **R-factors** | R(obs)= 6.37%, w$_R$(obs)= 6.27%, R(all)= 7.81%, w$_R$(all)=6.42% |
| **Bond Length(Å)** | Mn-O1=1.96846(8) <br> Mn-O2=2.05413(6) <br> Mn-O´2=2.00809(6) |
| **Bond Angle(°)** | Mn-O1-Mn= 144.7446(16) <br> Mn-O2-Mn= 154.8048(6) |

| Table 1(b): Atomic Parameters of polycrystalline GPMO ||||||||
|---|---|---|---|---|---|---|---|
| **Atoms** | **Atomic Positions** | | | **U$_{iso}$ factors** | **Occupancy** | **Wyckoff Position** | **Site Symmetry** |
| | **x** | **y** | **z** | | | | |
| **Gd** | 0.0779(5) | 0.25 | 0.9849(4) | 0.002 | 0.55 | 4c | m |
| **Pr** | 0.0779(5) | 0.25 | 0.9849(4) | 0.002 | 0.45 | 4c | m |
| **Mn** | 0 | 0 | 0.5 | 0.003 | 1 | 4b | -1 |
| **O1** | 0.5160(8) | 0.25 | 0.1096(1) | 0.036 | 1 | 4c | m |
| **O2** | 0.2165(9) | 0.5416(1) | 0.2039(9) | 0.011 | 1 | 8d | 1 |

**Table 2(a): Refined parameters and phase data of GPMO at different temperatures**

| Parameters/Temperature | 300 K | 200 K | 100 K | 50 K |
|---|---|---|---|---|
| a(Å) | 5.8401(5) | 5.8396(5) | 5.8379(5) | 5.8329(8) |
| b(Å) | 7.5125(6) | 7.5085(6) | 7.5025(6) | 7.4966(10) |
| c(Å) | 5.3765(4) | 5.3750(4) | 5.3727(4) | 5.3678(7) |
| **Volume(Å$^3$)** | **235.88(3)** | **235.67(3)** | **235.32(3)** | **234.72(5)** |
| GOF | 1.28 | 1.33 | 1.56 | 1.82 |
| Rp (%) | 7.47 | 7.70 | 8.46 | 7.20 |
| wRp(%) | 11.50 | 11.36 | 11.93 | 10.77 |
| R(obs) (%) | 4.82 | 4.32 | 4.78 | 4.10 |
| wR(obs) (%) | 10.68 | 5.46 | 6.71 | 4.94 |
| R(all) (%) | 5.03 | 4.32 | 4.81 | 4.50 |
| wR(all) (%) | 10.70 | 5.46 | 6.73 | 5.27 |

**Table 2(b): Atomic Parameters of GPMO at different temperatures**

| Atoms | 300 K | 200 K | 100 K | 50 K |
|---|---|---|---|---|
| **Gd/Pr** | | | | |
| x | 0.0738(2) | 0.0726(4) | 0.0721(4) | 0.0738(3) |
| y | 0.25 | 0.25 | 0.25 | 0.25 |
| z | 0.986(5) | 0.9850(4) | 0.9850(5) | 0.9847(4) |
| **Uiso factor** | 0.0403 | 0.0349 | 0.0492 | 0.0351 |
| **Mn** | | | | |
| x | 0 | 0 | 0 | 0 |
| y | 0 | 0 | 0 | 0 |
| z | 0.5 | 0.5 | 0.5 | 0.5 |
| **Uiso factor** | 0.0433 | 0.0325 | 0.0497 | 0.0315 |
| **O1** | | | | |
| x | 0.474(4) | 0.458(4) | 0.545(4) | 0.453(4) |
| y | 0.25 | 0.25 | 0.25 | 0.25 |
| z | 0.108(3) | 0.118(3) | 0.102(3) | 0.106(3) |
| Uiso factor | 0.0527 | 0.0574 | 0.0541 | 0.0897 |
| **O2** | | | | |
| x | 0.185(2) | 0.183(2) | 0.192(3) | 0.188(2) |
| y | 0.5583(14) | 0.5549(15) | 0.5561(15) | 0.5547(14) |
| z | 0.205(2) | 0.215(3) | 0.208(3) | 0.209(2) |
| Uiso factor | 0.0257 | 0.0054 | 0.033 | 0.0032 |
| **Bond length** | | | | |
| Mn-O1(Å) | 1.97157(15) | 1.99613(15) | 1.97409(11) | 1.9777(3) |
| Mn-O2(Å) | 1.96557(11) | 1.91459(11) | 1.97238(15) | 1.95087(19) |
| Mn-O2´(Å) | 2.19147(14) | 2.21874(14) | 2.15779(2) | 2.1783(3) |
| <Mn-O>(Å) | 2.04287 | 2.04315 | 2.0347 | 2.0356 |
| **Bond Angle** | | | | |
| Mn-O1-Mn(°) | 144.579(4) | 144.031(4) | 143.959(4) | 142.751(6) |
| Mn-O2-Mn(°) | 145.3543(17) | 147.4250(16) | 147.4815(17) | 147.370(3) |
| <Mn-O-Mn> (°) | 144.9666 | 145.728 | 145.720 | 145.060 |
| JT Distortion | 0.002647 | 0.003958(2) | 0.001828(3) | 0.0024846(5) |